\documentclass[aps,prb,10pt,reprint,groupedaddress,superscriptaddress]{revtex4-2}

\usepackage[usenames,dvipsnames]{xcolor}
\usepackage{amssymb}
\usepackage{graphicx}
\usepackage{amsmath}
\usepackage{txfonts}
\usepackage[bookmarks=true,colorlinks,citecolor=blue,urlcolor=blue]{hyperref}
\usepackage{braket}
\usepackage{babel}
\usepackage{blindtext}
\usepackage[compat=1.1.0]{tikz-feynman}
\usepackage[normalem]{ulem}
\newcommand{\cd}[1]{c^{\dagger}_{#1}}
\newcommand{\co}[1]{c^{}_{#1}}
\newcommand{\e}{\mathrm{e}}

\begin{document}

\newcommand{\OurTitle}{Hole Spectral Function of a Chiral Spin Liquid in the Triangular Lattice Hubbard Model}

\title{\OurTitle}

\newcommand{\TUM}{\affiliation{Department of Physics, Technical University of Munich, 85748 Garching, Germany}}
\newcommand{\MCQST}{\affiliation{Munich Center for Quantum Science and Technology (MCQST), Schellingstr. 4, 80799 M{\"u}nchen, Germany}}
\newcommand{\ugent}{\affiliation{Department of Physics and Astronomy, Ghent University, B-9000 Ghent, Belgium}}

\author{Wilhelm Kadow} \TUM \MCQST
\author{Laurens Vanderstraeten} \ugent
\author{Michael Knap} \TUM \MCQST

\begin{abstract}
Quantum spin liquids are fascinating phases of matter, hosting fractionalized spin excitations and unconventional long-range quantum entanglement. These exotic properties, however, also render their experimental characterization challenging and finding ways to diagnose quantum spin liquids is therefore a pertinent challenge. Here, we numerically compute the spectral function of a single hole doped into the half-filled Hubbard model on the triangular lattice using techniques based on matrix product states. At half filling the system has been proposed to realize a chiral spin liquid at intermediate interaction strength, surrounded by a magnetically ordered phase at strong interactions and a superconducting/metallic phase at weak interactions. We find that the spectra of these phases exhibit distinct signatures. By developing appropriate parton mean-field descriptions, we gain insight into the relevant low energy features. While the magnetic phase is characterized by a dressed hole moving through the ordered spin background, we find indications of spinon dynamics in the chiral spin liquid. Our results suggest that the hole spectral function, as measured by angle-resolved photoemission spectroscopy, provides a useful tool to characterize quantum spin liquids.
\end{abstract}

\maketitle

\section{Introduction}
\begin{figure}[b]
		\centering
		\includegraphics[width=0.48 \textwidth]{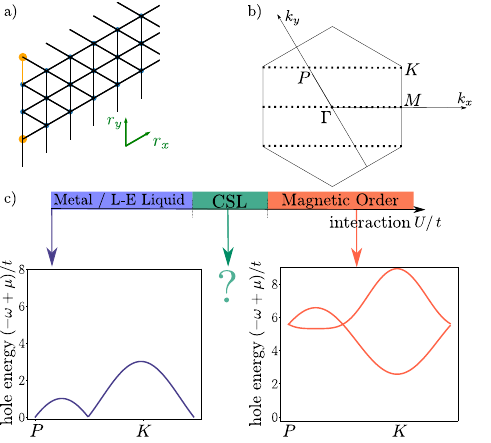}
		\caption{\textbf{Triangular lattice Hubbard model.} a) Our matrix product state calculations are performed on elongated cylinder geometries, with open boundary conditions in the $x$-direction and periodic ones in the $y$-direction (yellow sites are identified with each other). b) First Brillouin zone with dots indicating the allowed momenta for the lattice geometry shown in a). c) The proposed phase diagram of the triangular lattice Hubbard model on cylinder geometries \cite{Szasz2021} consists of a metal/Luther-Emery (L-E) liquid \cite{Raghu2010, Nandkishore2014, Gannot2020}, a chiral spin liquid (CSL), and a 120$^\circ$ magnetically ordered phase. The hole spectral functions are clearly distinct in the metallic and in the ordered phase. In this work, we show that the hole spectrum of the chiral spin liquid is distinct from the other phases.}
		\label{fig::lattice}
	\end{figure}

Quantum spin liquids are quantum disordered ground states that are characterized by fractionalized excitations and long-range topological entanglement~\cite{Savary2017, Knolle2019}. Powerful theoretical frameworks have been introduced to classify these exotic states of matter~\cite{Wen1989a, Wen1990, Wen1991, Wen2007} and exactly solvable models provide crucial insights into their fundamental properties~\cite{Kitaev2003, Kitaev2006}; aspects of which have recently been studied on quantum devices~\cite{Satzinger2021, Semeghini2021}. Moreover, numerical approaches based on tensor networks as well as variational Monte Carlo simulations suggest that quantum spin liquids are stabilized in various models. In particular, recent matrix product state (MPS) calculations on cylinder geometries indicate that the Hubbard model on the triangular lattice realizes at half-filling a chiral spin liquid for intermediate interaction strength ~\cite{Szasz2020, Chen2021, Szasz2021, Cookmeyer2021}. The chiral spin liquid~\cite{Kalmeyer1987, Wen1989} is surrounded by a long-range ordered phase with a 120$^\circ$ spiral order at strong interactions~\cite{Huse1988,White2007} and a possibly superconducting or metallic phase at weak interactions~\cite{Raghu2010, Nandkishore2014, Gannot2020, Szasz2020}; see Fig.~\ref{fig::lattice}c) for a sketch of the phase diagram. 

Triangular lattice systems have several potential experimental realizations, including organic compounds~\cite{Shimizu2003, Itou2008}, transition metal dichalcogenides~\cite{Law2017, Ruan2021, Tang2020}, and triangular lattice materials~\cite{Shen2018, Scheie2021}.
However, due to the absence of any conventional order, it is challenging to experimentally detect quantum spin liquids. Spectroscopic probes offer a promising route to identify characteristic signatures of quantum spin liquids at finite frequencies. Among them is inelastic neutron scattering, often leading to spectral features that are smeared out over the Brillouin zone at comparatively high energies, \textit{c.f.}~\cite{Knolle2014, Punk2014}. Spin-polarized scanning tunneling microscopy~\cite{Feldmeier2020, Konig2020} and nitrogen-vacancy (NV) -center magnetometry~\cite{Klocke2021} can provide an alternative route to probe charge neutral edge states in chiral spin liquids. All these dynamical probes measure in some form the dynamical structure factor, which creates spin-1 excitations of the underlying state. However, since the fundamental excitations of quantum spin liquids are fractionalized, it seems pertinent to investigate probes that offer a more direct view on spinons that carry spin~1/2~\cite{Lauchli2004, Bohrdt2020}.

In this work, we compute the spectral function of a single hole injected in the ground state of the triangular lattice Hubbard model at half-filling. The hole excitation carries spin 1/2 and charge, and its spectrum is directly measured with angle-resolved photoemission spectroscopy (ARPES). Recent theoretical work has shown that ARPES can be used to probe fractional excitations in square-lattice antiferromagnets~\cite{Bohrdt2018, Bohrdt2020} and in frustrated quantum magnets~\cite{Lauchli2004}. We compute the spectral function numerically with algorithms based on MPS, and we find that the hole spectral functions are distinct in the different phases of the triangular lattice Hubbard model. We furthermore develop parton mean-field descriptions for the relevant low-energy excitations in the different phases. Our findings suggest that ARPES can be a useful probe for characterizing quantum spin liquids.

\section{Model and method}
We consider the Hubbard model on the triangular lattice,
    	\begin{equation}\label{eq::Hubbard}
    	H = -t \sum_{\substack{\left\langle i,j \right\rangle \\
    	\sigma=\uparrow,\downarrow}} \left(\cd{i\sigma} \co{j\sigma} +\mathrm{h.c.} \right) +U \sum_{j} n_{j\uparrow}n_{j\downarrow} ,
    	\end{equation}
	where $t$ is the hopping matrix element and $U$ the on-site interaction strength. Fermions with spin $\sigma$ are created/annihilated by $\cd{j\sigma}$/$\co{j\sigma}$, and $n_{j\sigma}=\cd{j\sigma} \co{j\sigma}$ are the corresponding number operators.
	
	We compute the optimal MPS representation of the ground state $\ket{\psi_0}$ of the triangular lattice Hubbard model at half filling with DMRG (using the library TeNPy~\cite{Hauschild2018}) on elongated cylinders, depicted in Fig.~\ref{fig::lattice}a). Open boundary conditions are assumed in the $x$-direction and periodic ones in the $y$-direction.
	
	We compute the hole spectral function by time-dependent MPS simulations,
	\begin{equation}
	    A(\mathbf{k},\omega) = \sum_{\sigma=\uparrow,\downarrow} \int \mathrm{d} \tau \e^{i\tau \omega} \bra{\psi_0}\cd{\mathbf{k}\sigma}(\tau) \co{\mathbf{k}\sigma}(0) \ket{\psi_0}.
	    \label{eq::A}
	\end{equation}
	The maximum time until which we can accurately simulate the evolution with MPS methods is limited by the growth of entanglement, which leads to a finite frequency resolution in the spectral function. Here, we use a variation on established methods \cite{Zaletel2014, Paeckel2019}: instead of evolving the action of a local operator, we time-evolve a plane wave with fixed momentum $\mathbf{k}$ on a finite cylinder. Following hydrodynamic arguments, finite momentum excitations will decay exponentially at late times (with a decay constant proportional to $|\mathbf{k}|^2$), leading to a reduced entanglement growth.
	As a consequence, for a given bond dimension we can evolve the state to longer times and thus obtain better resolution of the spectral function. This intuition is also supported by direct numerical comparisons in the appendix, in which also further details on the method are presented; see Appendix~\ref{app::method_details}. We furthermore implement $\mathrm{U}(1)\times \mathrm{U}(1)$ symmetries for particle number and spin conservation respectively as well as the momentum conservation in the $y$-direction \cite{Motruk2016,Ehlers2017}. This allows us to use bond dimensions of up to $\chi=1500$ for the time evolved state. Despite these efforts, the MPS simulations are challenging for these two-dimensional systems, since the numerical cost grows exponentially with the cylinder circumference. In order to keep the truncation error small for the bond dimension stated above, we restrict ourselves to cylinders with $L_y =3$. The Brillouin zone associated with this lattice geometry is sketched in Fig.~\ref{fig::lattice}b). 
	
	We complement these spectral function calculations by a variational computation of the lowest-lying excited states. In this method, we start from an MPS approximation for the ground state on the infinite cylinder, and optimize a variational ansatz that models the excited states on top of the ground-state MPS with fixed momentum. This variational approach, which can be interpreted as the MPS version of the single-mode approximation, was introduced for computing the spectrum of spin chains \cite{Haegeman2012} and recently extended to cylindrical geometries \cite{VanDamme2021}.

	\begin{figure}[t]
		\centering
		\includegraphics[width=0.48 \textwidth]{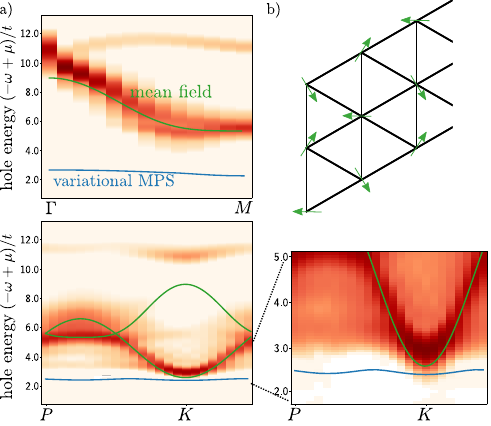}
		\caption{\textbf{Hole spectral function in the ordered phase.} a) Spectral function $A(\mathbf{k},\omega)$ for $U=12.0\,t$ along the two distinct cuts of the cylinder. The most prominent branch is described by the mean-field dispersion (green line). The full spectral function is compared with a variational MPS calculation, which picks up the lower edge of the spectrum which for most momenta carries only little spectral weight (\textit{c.f.} magnified data on the right hand side). 
		b) Illustration of the long-range ordered state with 120$^\circ$ spiral order.}
		\label{fig::spectrumU12}
	\end{figure}

\section{Magnetically ordered phase}
Using this algorithm, we compute the hole spectral function in the long-range ordered phase for strong interactions $U=12.0\,t$; see Fig.~\ref{fig::spectrumU12}. The excitation energies of the hole are shifted by a chemical potential $\mu$, chosen such that both particle and hole excitation energies are equal. This allows us to directly read off half of the charge gap from the smallest excitation energy in the spectral function. The spectrum is dominated by a single branch, which for all wave vectors carries most of the spectral weight. 

To understand the origin of the well-defined excitation branch, we consider a mean-field description of a hole on a triangular lattice which experiences an effective magnetic field arising from the long-range 120$^\circ$ spiral order of the ground state,
	\begin{equation}\label{eq::magnetic_mf}
	 H^\text{mf}= -t \sum_{\langle i,j \rangle, \sigma} \left(\cd{i\sigma} \co{j\sigma} + \mathrm{h.c.}\right) + h \sum_{i} \mathbf{M_i}\cdot\mathbf{S_i}.
	\end{equation}
Here, $t$ is the hopping matrix element and $h$ is the effective magnetic field. The spin operators are defined as $\mathbf{S_i}=\frac{1}{2} (\cd{i\uparrow}, \cd{i\downarrow}) \boldsymbol{\sigma} (\co{i\uparrow}, \co{i\downarrow})^T$ where $\boldsymbol{\sigma}$ is the vector of Pauli matrices. The 120$^\circ$ magnetic order is set by  $\mathbf{M_i}=(\cos(\mathbf{Q\cdot R_i}), \sin(\mathbf{Q\cdot R_i}), 0)$ with $\mathbf{Q}=(0,\frac{4\pi}{3})$, as sketched in Fig.~\ref{fig::spectrumU12}b). Such a Hamiltonian is also regularly used as a starting point for variational Monte Carlo simulations to describe the magnetically ordered phase on the triangular lattice \cite{Iqbal2016, Ferrari2019,Tocchio2020, Tocchio2021}. By diagonalizing the Hamiltonian, we obtain the effective hole dispersion
	\begin{equation}\label{eq::magnetic_dispersion}
	\omega_\mathbf{k}^\text{mf} = \frac{1}{2} \left(\epsilon_\mathbf{k} +\epsilon_{\mathbf{k}\pm\mathbf{Q}} + \sqrt{(\epsilon_\mathbf{k} -\epsilon_{\mathbf{k}\pm\mathbf{Q}})^2+h^2} \right),
	\end{equation}
where $\epsilon_\mathbf{k}$ is the single-particle dispersion on the triangular lattice with a band width that is proportional to the hopping strength $t$.
	
To quantitatively compare the mean-field dispersion $\omega_\mathbf{k}^\text{mf}$ with our spectral function obtained from time-evolving the MPS, we identify the effective field strength $h$ with its mean-field value $h=Um$, where $m$ is the magnetic order parameter $m=\langle \mathbf{M_i}\cdot\mathbf{S_i} \rangle$ of the MPS ground state. To determine $m$, we explicitly have to break the $\mathrm{SU}(2)$ symmetry with a small pinning field on every third lattice site~\cite{White2007}. With this identification of the effective magnetic field $h$, the mean-field ansatz quantitatively reproduces the dominant branch of the hole spectrum without any additional fit parameters, see Fig.~\ref{fig::spectrumU12}a). 
	
The dominant branch of the spectral function in the magnetically ordered phase can hence be understood as a hole moving in an effective field, which is generated by the spin background. Apart from the main branch, continua are visible in particular for the $P$-$K$ cut, indicating the importance of many-body excitations. Moreover, a rather well defined high-energy branch exists as well.
	
These results are supported by the variational MPS calculations which find the lowest energy excitations for a given momentum and fixed quantum numbers, irrespective of their spectral weight. As a consequence, we expect the results from the variational ansatz to coincide with the lower edge of the full spectrum. As shown in the zoom in Fig.~\ref{fig::spectrumU12}a), there is a good agreement at the $P$-$K$ momentum cut between the different approaches. However, at the $\Gamma$-$M$ cut the variational MPS dispersion differs strongly because most of the spectral weight is carried by the mean-field branch. This observation is confirmed by calculating the spectral weight directly from the variational excited-state wave function, which is very small for all momenta except at the K-point. There the variational dispersion touches the dominant spectral branch.
Furthermore, our results are in good agreement with recent calculations based on the self-consistent Born approximation \cite{Chen2022}.
	
	\begin{figure}[t]
		\centering
		\includegraphics[width=0.48 \textwidth]{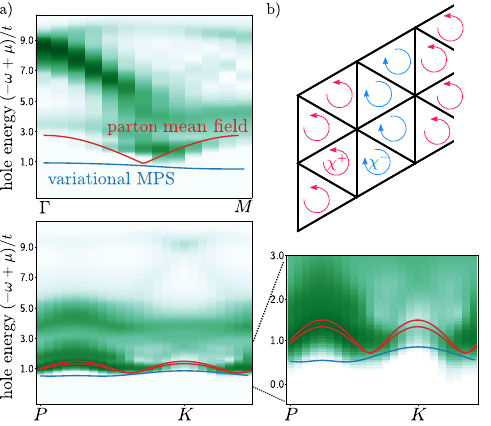}
		\caption{\textbf{Hole spectral function in the chiral spin liquid.} a) Spectral function $A(\mathbf{k},\omega)$ for $U=7.6\,t$ along the two distinct cuts of the cylinder. The parton mean-field theory captures the dominant spectral response at low energies along both cuts. Variational MPS again picks up the lower edge of the spectrum. 
		b) Sign pattern of the chiral order parameter on our finite size cylinders where $|\chi^+| \neq |\chi^-|$. The unit cell consists of two sites, leading to two bands in the parton mean-field theory.}
		\label{fig::spectrumU76}
	\end{figure}

\section{Chiral spin liquid phase}
We now turn our attention to the chiral spin liquid which was found to occur at intermediate interactions \cite{Szasz2020}. Specifically, we choose an interaction strength of $U=7.6\,t$, in the middle of the small parameter range for the intermediate phase. It is apparent in Fig.~\ref{fig::spectrumU76}a) that the spectrum at low energies looks qualitatively very different from the magnetically ordered phase. The spectral weight is not concentrated in one main branch anymore but rather is broadened over a wide range of energies with some noticeable features. The dispersion at low energies has drastically changed: While in the magnetically ordered phase, the minimal excitation energy is at the $K$ point, in the chiral spin liquid the lower edge of the spectrum has a maximum at this momentum. 

A strong indication for the chiral spin liquid, which breaks time-reversal symmetry spontaneously, is a non-vanishing order parameter $\chi_{ijk} = S_i \cdot (S_j\times S_k)$ around triangles in the ground state. We indeed find similar values for the order parameter as in Ref.~\cite{Szasz2020} and also observe the same pattern of triangles with different magnitude and opposing signs of $\chi$, Fig.~\ref{fig::spectrumU76}b). However, this particular pattern is a distinct aspect of the three-leg cylinder~\cite{Szasz2021}.

To gain an understanding of the hole dynamics we compare our MPS results to a parton mean-field ansatz. In Ref.~\cite{Song2021} the authors present a spinon mean-field Hamiltonian for a hole doped $\mathrm{U}(1)$ chiral spin liquid as well as an ansatz for a projected $d+id$ superconductor. Here we modify this mean-field approach to account for the particular flux pattern on the three-leg cylinder. We start from the mean-field Hamiltonian
	 \begin{equation}\label{eq::chiral_mf}
	 H^\text{mf}_\text{parton} = \sum_{\langle i,j \rangle, \sigma} \left( J_{ij} \, f^\dagger_{i\sigma} f^{}_{j\sigma} + \mathrm{h.c.}\right),
	 \end{equation}
where $f^\dagger_{i\sigma}$ ($f^{}_{i\sigma}$) create (annihilate) spinons at low energies. The hopping terms are chosen with equal amplitude but different phases $J_{ij}=Je^{i\theta_{ij}}$, such that the sum of $\theta_{ij}$ around each triangle leads to the pattern shown in Fig.~\ref{fig::spectrumU76}~b). We now determine the values of $\theta_{ij}$ from the phase of $\langle \cd{i\sigma} \co{j\sigma} \rangle$ in the MPS representation of the three-leg cylinder ground state. This is motivated by performing a Schrieffer-Wolf transformation, in which to leading order the spinons couple to the dynamically generated gauge field in the effective spin Hamiltonian \cite{Motrunich2005}. Therefore, the spinons directly inherit the flux penetrating through the triangles.
When setting the amplitude of the hopping to the superexchange energy, which is the typical energy scale for spinons, $J=4(1-7t^2/U^2)t^2/U$~\cite{Cookmeyer2021}, we find a remarkable fit to the dispersion of the lower spectral edge, again without additional fitting parameters; Fig.~\ref{fig::spectrumU76}~a). Therefore, we have strong indications that the dominant low energy physics is actually determined by spin dynamics. The alternating sign of the chiral order parameter effectively leads to a doubling of the unit cell, which is directly reflected both in the numerical and in the mean-field spectrum. Crucially, the mean-field Hamiltonian not only captures the shape of the dispersion quite well, but also predicts the distribution of spectral weight properly, see Appendix~\ref{app::chiral_mf}. In the appendix we also demonstrate that our modified, microscopic mean-field ansatz indeed shows better agreement than the generic approach of Ref.~\cite{Song2021}, which may be suitable for larger systems and hence also for experimental realizations.

For this phase, the excitations found by the variational MPS approach again provide a lower bound on the full spectrum.
Similar to the magnetic phase, the deviations could be caused by complex many-body excitations with very low spectral weight. Nevertheless, the variational results coincide approximately with the low-energy edge of the full spectrum for the $P$-$K$ cut. With both methods we find two humps and in particular a local maximum of the dispersion at the $K$ point, in stark contrast to the magnetically ordered phase.

\section{Discussions \& Outlook}
Our results demonstrate that the hole spectral function of the chiral spin liquid in the triangular lattice Hubbard model is clearly distinct from the one in competing phases. Doping a hole in the ground state can be interpreted as simultaneously creating a spinless holon and a spinon that carries spin-1/2. A hole excitation thus has the potential to offer a more direct view on fractionalized excitations.
Our parton mean-field description of an excitation moving in the chiral order parameter field captures key features of the low-energy spectrum. In particular, the dispersion of the lower spectral edge is set by the superexchange energy scale, which suggests dynamics originating from spinons rather than dressed holes.

For future studies, it would be interesting to investigate the effective interaction between the holon and the spinon. At finite doping, holons can condense and spinons determine the low energy properties of the spectrum \cite{Jiang2020, Zhu2020, Song2021, Peng2021}. Similar spectral signatures are expected for a single spinon deconfined from the holon. Alternatively, there could in principle be a holon-spinon bound state, as on the square lattice antiferromagnet \cite{Bohrdt2020, Bohrdt2020a}. We have investigated the possibility of a holon-spinon binding via geometric strings \cite{Grusdt2018, Grusdt2019} on the triangular lattice. However, since the local correlations are much smaller there than on the square lattice, the minimum size of the bound object, if it exists, exceeds by far the circumference of our lattice. Hence, a spinon-holon binding via geometric strings is unlikely for our small system but could be interesting to explore in future work.
Furthermore, a better understanding of the interaction between the constituting partons would help to gain further insights in the high energy features of the spectra, certain aspects of which seem similar in the distinct phases.

Our results suggest that dynamical hole spectra, as experimentally measured by ARPES, are potentially a very useful probe to characterize the low-energy dynamics of spin-liquid materials and, in particular, analyze the fractionalization of the elementary excitations. Given the recent progress on finding new candidate materials, especially the triangular-lattice systems, this avenue opens up the exciting possibility of experimentally observing spin-liquid physics in a more direct way.

The MPS code and data analysis are available on Zenodo upon reasonable request \cite{Zenodo}

\section*{Acknowledgements}
We thank A. Bohrdt, M. Drescher, F. Grusdt, J. Motruk, and F. Pollmann for insightful discussions.
We acknowledge support from the Deutsche Forschungsgemeinschaft (DFG, German Research Foundation) under Germany’s Excellence Strategy--EXC--2111--390814868, TRR80 and DFG grants No. KN1254/1-2 and No. KN1254/2-1, BMBF EQUAHUMO, the European Research Council (ERC) under the European Union’s Horizon 2020 research and innovation programme (Grant Agreement No. 851161), as well as the Munich Quantum Valley, which is supported by the Bavarian state government with funds from the Hightech Agenda Bayern Plus. LV is suported by the Research Foundation Flanders.

\appendix
\section{Details on the Method}\label{app::method_details}
 
    Here we give additional details on the MPS method for evaluating the spectral function. Our simulations were performed with the tensor network library TeNPy \cite{Hauschild2018}. We have used $\mathrm{U}(1)\times \mathrm{U}(1)$ symmetries for particle number and spin conservation respectively. Additionally, we work in a mixed real and momentum space to impose momentum conservation around the cylinder \cite{Motruk2016, Ehlers2017}, gaining an extra $\mathbb{Z}_3$ symmetry. The time evolution is obtained from an MPO representation of the time evolution operator (the $W_{\mathrm{II}}$ operator \cite{Zaletel2014}) with a variational truncation scheme with fixed MPS bond dimension and step size $\delta\tau = 0.025/\,t$.
	
    For evaluating the spectral function we have used a variation of the standard approaches in the MPS literature \cite{Paeckel2019}. Instead of time-evolving an initial state with a local excitation, we use a plane wave with a fixed momentum, which is common in other MPS approaches for computing spectral functions \cite{Kuhner1999, Dargel2011, Holzner2011}. We first determine an MPS approximation for the ground state at half filling $\ket{\psi_0}$ using the DMRG algorithm on a finite $L_y\times L_x$ system, with $L_y=3$ and $L_x=24$, chosen such that boundary effects do not play a role. On top of this state, the time dependent correlation function is given by 
	\begin{equation}\label{eq::Cij}
		C_{ij}^{\sigma}(\tau) = \braket{\psi_0| \e^{i \tau H} \cd{j\sigma} \e^{-i \tau H} \co{i\sigma}|\psi_0}.
	\end{equation}
    We assume that $\ket{\psi_0}$ is a good approximation to the ground state with energy $E_0$. We indeed confirm this by time evolving $\e^{-i \tau H}\ket{\psi_0}$, which yields $\e^{-i \tau (E_0+\delta\epsilon)}\ket{\psi_0}$. The small error $\delta\epsilon$ from the time evolution method is then corrected for the spectral function. Therefore Eq.~\ref{eq::Cij} reduces to
	\begin{equation}
		\tilde{C}_{ij}^\sigma(\tau) = \braket{\psi_0|\cd{j\sigma} \e^{-i \tau H} \co{i\sigma}|\psi_0}.
	\end{equation}
    This quantity is usually evaluated by time evolving the state after hole creation at the origin, and then taking the overlap with the states where a hole is inserted at site $\mathbf{j}$. As explained in the main text in the case of $U=7.6\,t$ our MPS approximation to the ground state explicitly breaks translational symmetry in $r_x$-direction such that we also have to compute the time evolution after hole creation at the site next to the origin.
	
    This procedure is limited by the time for which we can feasibly represent the state as an MPS with a certain bond dimension, as the entanglement entropy in the time-evolved state is expected to grow linearly in time.
	
    However, we are actually interested in the spatial Fourier transform of this quantity,
	\begin{equation}
	\tilde{A}^\sigma(\mathbf{k},\,\tau) = \sum_j \e^{-i \mathbf{k}\cdot\mathbf{r}_j} \tilde{C}^\sigma_{0j}(\tau).
	\end{equation}
    Therefore, instead of computing the time evolution for the initial state where a single hole is created locally at the origin, we can also perform a time evolution of a state with the hole inserted as a plane wave,
	\begin{equation} \label{eq:tildeA}
	\tilde{A}^\sigma(\mathbf{k},\tau) = \bigg( \bra{\psi_0} \sum_{j} \e^{-i\mathbf{k}\cdot\mathbf{r}_j}\cd{j\sigma}       \bigg) \e^{-iH\tau} \bigg( \co{0\sigma}\ket{\psi_0} \bigg).
	\end{equation}
    
\begin{figure}[h]
		\centering
		\includegraphics[width=0.45 \textwidth]{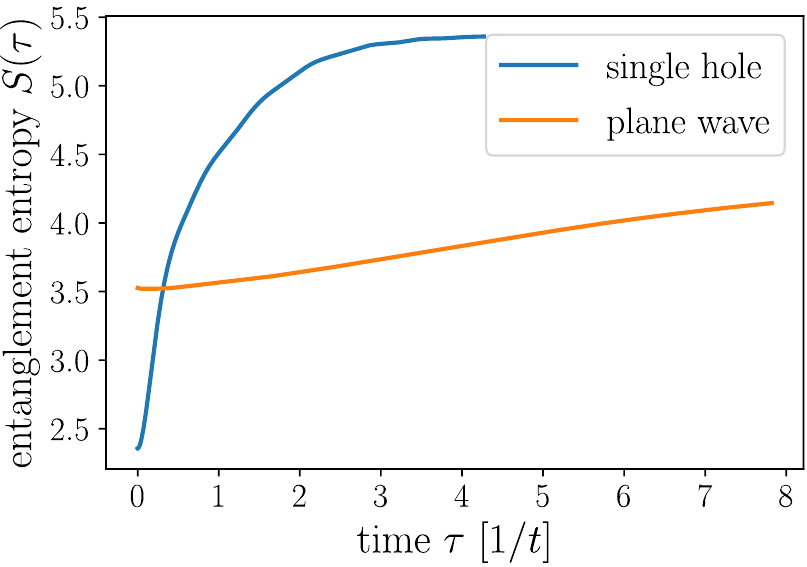}
		\caption{\textbf{Growth of entanglement entropy.} We compare the growth of the half-cylinder entanglement entropy after inserting a single hole locally in the ground state for $U=12\,t$ and after inserting a hole as a plane wave with momentum $k_x = 0$.}
		\label{fig::entropy}
    \end{figure}

    By evolving the momentum superposition (bra-vector) instead of a local-hole state (ket-vector), we can hope that the entanglement growth is smaller, because correlations decay faster. Indeed, in Fig.~\ref{fig::entropy} we observe that although the initial entanglement of the plane wave is larger (due to the fact that a plane wave is a superposition of MPS), it acquires much less entanglement over time.

    A disadvantage of the plane waves is that we have to do a separate time evolution for each allowed value of the momentum. However, this process can be easily parallelized. Finally, note that the time evolution operator in Eq.~\eqref{eq:tildeA} can be split into $\tau=\tau_1+\tau_2$, to perform two separate time evolutions, one on the bra-vector and one on the ket-vector, which further increases the latest time to feasibly represent the MPS.

    \begin{figure}[t]
		\centering
		\includegraphics[width=0.45 \textwidth]{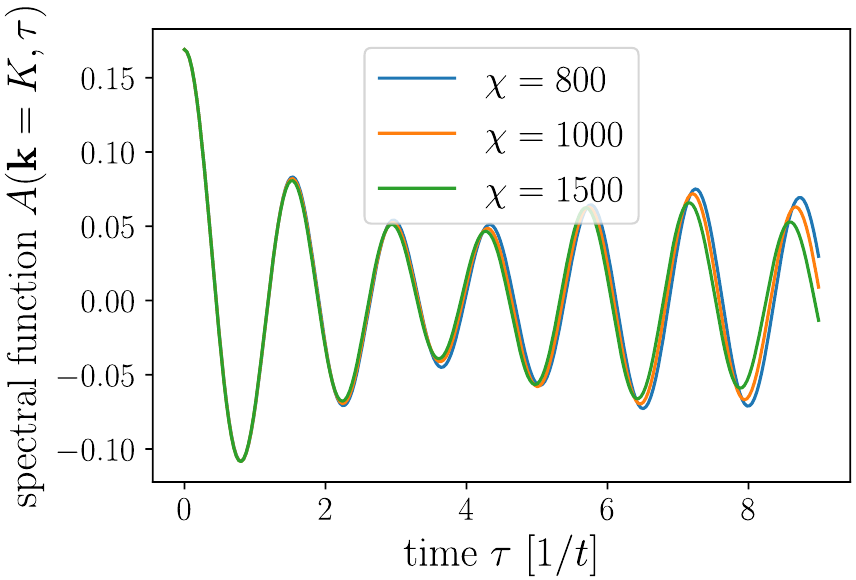}
		\caption{\textbf{Convergence of spectral function.} The spectral function in time at the $K$ point is converged for different values of the maximal MPS bond dimension $\chi$ and only small deviations arise at late times.}
		\label{fig::chi_compare}
	\end{figure}

	Although the plane waves allow for evolution to longer times, we are still restricted to maximal times of order $\tau_{\mathrm{max}}\approx 10$, which will yield a finite frequency resolution for the spectral function.
	Hence, we use linear prediction \cite{White2008} combined with a Gaussian envelope before taking the temporal Fourier transform	
	\begin{equation}
	    A(\mathbf{k},\omega) = \sum_\sigma \int \mathrm{d} \tau \e^{i\tau (E_0+\omega)} \tilde{A}^{\sigma}(\mathbf{k},\tau).
	\end{equation}
	
	To ensure that the time evolution is indeed well converged for late times, we compare the results for MPS with different bond dimensions in Fig.~\ref{fig::chi_compare}. For small times $A(\mathbf{k},\tau)$ does not depend on the value of $\chi$. However, for longer time evolutions we can see deviations between the various bond dimensions. Nevertheless, the qualitative behavior remains the same.

\section{Parton Mean-Field Theory for the Chiral Spin Liquid}\label{app::chiral_mf}

    As discussed in the main text, we construct a parton mean-field Hamiltonian to understand the low-energy features of the MPS hole spectrum in the chiral spin liquid phase at $U=7.6\,t$,
    \begin{equation}\label{eq::chiral_mf2}
	 H^\text{mf}_\text{parton} = \sum_{\langle i,j \rangle, \sigma} \left( J e^{i\theta_{ij}} \, f^\dagger_{i\sigma} f^{}_{j\sigma} + \mathrm{h.c.}\right).
	\end{equation}
	This ansatz is motivated by the $\mathrm{U}(1)_2$ chiral spin liquid that was introduced by Kalmeyer-Laughlin~\cite{Kalmeyer1987} and recently investigated in the context of hole doping~\cite{Song2021}. There the phases $\theta_{ij}$ were taken to incorporate a $\pi/2 \pm 3\theta$ flux through up/ down pointing triangles.
	
\begin{figure}[t]
		\centering
		\includegraphics[width=0.5 \textwidth]{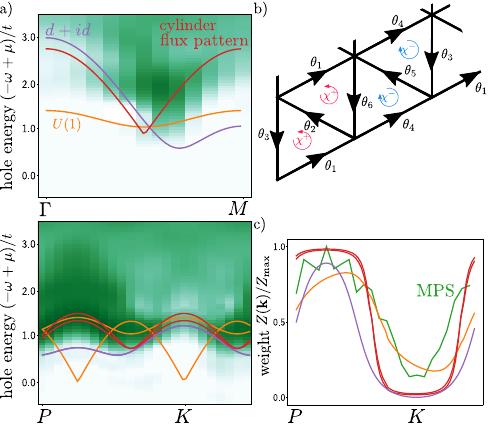}
		\caption{\textbf{Comparison of mean-field theories.} a) We compare the MPS spectrum to the $\mathrm{U}(1)$ chiral spin liquid and the projected $d+id$ superconductor from~\cite{Song2021} as well as our flux pattern ansatz on the cylinder. b) The doubled unit cell and corresponding hopping phases used in Eq. \eqref{eq::chiral_mf2}. c) The spectral weight of the mean-field approaches compared to the MPS data. For the $\mathrm{U}(1)$ ansatz the weight is only shown for the band that fits the MPS spectrum.}
		\label{fig::chiral_mf}
	\end{figure}

	We also introduce a modification of the $\mathrm{U}(1)_2$ chiral spin liquid. Instead of the $\pi/2 \pm 3\theta$ flux, we follow the MPS results more closely and choose the $\theta_{ij}$ to reproduce the same chiral order sign pattern as measured on the cylinder in the ground state. Imposing this pattern to the flux we are still left with four free parameters $\theta_i$ for the phases around the doubled unit cell shown in Fig.~\ref{fig::chiral_mf}b), which we determine directly from the corresponding expectation values in the MPS ground state:
	\begin{align*}
	&\theta_1 \approx 0.004 \quad \theta_2 \approx 0.004 \\
	&\theta_3 \approx 0.322 \quad \theta_4 \approx -0.001 \\
	&\theta_5 = \theta_1 + \theta_2 - \theta_4 \\
	&\theta_6 = -2\theta_1 - 2\theta_2 -\theta_3 
	\end{align*}
	The equations for $\theta_5$ and $\theta_6$ ensure the correct pattern for the mean-field Hamiltonian, where the fluxes can only differ in sign but not in absolute value. Since the MPS has $|\chi^+|~\neq~|\chi^-|$ the measured values of $\theta_5$ and $\theta_6$ vary slightly.
	
	Besides the $\mathrm{U}(1)_2$ chiral spin liquid and our modified flux pattern ansatz, we compare the spectrum to a projected $d+id$ superconductor,
	\begin{equation}
	    H^{d+id}_\text{parton} = \sum_{k,\sigma} \epsilon_k f^\dagger_{k\sigma} f^{}_{k\sigma} + \sum_k \Delta_k \left( f^\dagger_{k\uparrow} f^\dagger_{-k\downarrow} + \mathrm{h.c.} \right) ,
	\end{equation}
	where $\epsilon_\mathbf{k}$ is the single-particle dispersion on the triangular lattice and with $\omega=e^{i2\pi/3}$ the pairing is given by
	\begin{equation}
	    \Delta_k = 2\Delta\left[\cos(\mathbf{k\cdot r_x})+ \omega  \cos(\mathbf{k\cdot r_y}) +\omega^2\cos(\mathbf{k}\cdot(\mathbf{r_y}-\mathbf{r_x} ))\right].
	\end{equation}
    This ansatz is used as a starting point for variational Monte Carlo simulations~\cite{Tocchio2020, Tocchio2021}, and hole doping was studied for this ansatz in Ref.~\cite{Song2021}. 

	In Fig.~\ref{fig::chiral_mf}a), we show the dispersion of the three different mean-field theories compared to the MPS spectrum. At the $\Gamma$-$M$ cut, there is only very little weight at low energies. Therefore, we concentrate on the $P$-$K$ and adjust all free parameters of the U$(1)_2$ chiral spin liquid and the $d+id$ superconductor to fit the well defined spectral edge. 
	
	In our flux pattern ansatz, shown in the main text, all flux parameters are determined by MPS expectation values. Thus the only adjustable variable is the hopping constant $J$. We find very good agreement when setting this to the superexchange energy $J=4(1-7t^2/U^2)t^2/U$~\cite{Cookmeyer2021}.
	Note that the flux pattern mean-field Hamiltonian has a doubled unit cell and hence we get two different bands, which overlap on the $\Gamma$-$M$ cut but slightly differ for the $P$-$K$ cut. 
	
	The U$(1)_2$ ansatz has a doubled unit cell as well. 
	However, it was not possible to fit both bands to the full spectrum. One of the two branches fits well to our data when selecting $\theta$ in the vicinity of $\theta=\pm\pi/6$, at which the ansatz describes a Dirac spin liquid. However, the second branch hosting the Dirac cones is absent in the numerical spectrum. Setting the hopping constant to the superexchange energy gives the best agreement for the bandwidth.
	
	When comparing to the $d+id$ superconductor, we can not only adjust the hopping constant, but have also the free variable $\Delta$, which is optimized to fit the low energy spectrum in the $P$-$K$ cut at around $\Delta \approx 0.4\,J$. The bandwidth is again determined by the superexchange scale. 
	
    In summary, for all three mean-field theories we find $J=4(1-7t^2/U^2)t^2/U$ indicating that the basic constituents are spinons rather than (dressed) holes.
	Moreover, we compare the spectral weight $Z(\mathbf{k})$ predicted by the mean-field approaches to the MPS results, Fig.~\ref{fig::chiral_mf}c). To extract the weight from the MPS spectrum, we fit the height and width of a Gaussian to the first peak. We find that the weight distribution is similar for all three mean-field approaches.

\end{document}